\documentclass[12pt]{article}
\usepackage{amssymb}
\renewcommand{\today}{September 1996}

\begin{document}
\sloppy
\begin{titlepage}
\null
\vspace{5mm}
\begin{flushright}
\begin{tabular}{l}
DFTT 62/96\\
hep-ph/9610291\\
\today
\end{tabular}
\end{flushright}
\vfill
\begin{center}
\Large
\textbf{Inclusive neutrino and antineutrino processes
and the problem of the spin of proton}\footnote
{\normalsize
Talk presented by S.M. Bilenky at the
{\it
$12^{\mathrm{th}}$
International Symposium on High Energy
Spin Physics},
SPIN96,
Amsterdam, September 10-14, 1996.}
\\[5mm]
\normalsize
W.M. Alberico$^{\mathrm{a}}$,
S.M. Bilenky$^{\mathrm{a,b}}$
and
C. Giunti$^{\mathrm{a}}$
\\[3mm]
(a) Dipartimento di Fisica Teorica,
Universit\`a di Torino and INFN,
Sezione di Torino,\\
Via P. Giuria 1, I-10125 Torino, Italy
\\
(b) Joint Institute for Nuclear Research, Dubna, Russia
\end{center}
\vfill
\begin{center}
\textbf{Abstract}
\\[3mm]
\begin{minipage}{0.8\textwidth}
The inclusive NC and CC neutrino(antineutrino)-nucleus scattering
for the case of a nucleus
with isotopic spin equal to zero is considered.
It is shown that the measurement of
the neutrino-antineutrino asymmetry
at relatively small energies could allow to obtain
model-independent
information about the contribution of
the axial strange current
to the cross sections of the NC processes.
\end{minipage}
\end{center}
\vfill
\null
\end{titlepage}

Recent experiments on the deep inelastic scattering
of longitudinally polarized leptons on longitudinally
polarized nucleons
\cite{recent_experiments}
confirmed the important EMC
\cite{EMC}
result:
the one-nucleon matrix element of the axial strange current
is relatively large and comparable with the matrix elements of
the axial $u$ and $d$ currents.
From a recent analysis of the data it follows that
\cite{Ellis-Karliner}
\begin{equation}
\begin{array}{l} \displaystyle
\Delta u
=
0.82 \pm 0.03
\;,
\\ \displaystyle
\Delta d
=
- 0.44 \pm 0.03
\;,
\\ \displaystyle
\Delta s
=
- 0.11 \pm 0.03
\;.
\end{array}
\label{01}
\end{equation}
The constants
$ \Delta q $
(with $q=u,d,s$)
are determined by
\begin{equation}
\left\langle
p \, , \, r
\left|
\,
\bar{q} \, \gamma^{\alpha} \, \gamma_{5} \, q
\,
\right|
p \, , \, r
\right\rangle
=
2 M
\,
r
\,
s^{\alpha}
\,
\Delta q
\;,
\label{02}
\end{equation}
where
$
\left|
p \, , \, r
\right\rangle
$
is the state of a nucleon with
momentum $p$ and projection of the spin on the direction
$s^{\alpha}$
equal to $r$
and $M$
is the mass of the nucleon.

The usual determination
of the constants
$ \Delta q $
from the data
on deep inelastic 
lepton-nucleon scattering and
from other data requires,
however,
several assumptions.
Probably the most
questionable assumption
is about the Regge behavior of the polarized
structure function
$g_1$
at small x.

As it is well known
(see Ref.\cite{Kaplan-Manohar}),
the investigation of NC-induced processes
allow us to obtain information about the matrix elements of
the axial and vector strange currents in a direct way.
The neutral current of the standard model in
the $u$, $d$, $s$ approximation has the form
\begin{equation}
j^Z_{\alpha}
=
v^{3}_{\alpha}
-
2 \sin^2\theta_{W}
\,
j^{\mathrm{em}}_{\alpha}
+
a^{3}_{\alpha}
-
\frac{1}{2}
\,
v^{S}_{\alpha}
-
\frac{1}{2}
\,
a^{S}_{\alpha}
\;,
\label{03}
\end{equation}
where
$v^{3}_{\alpha}$
and
$a^{3}_{\alpha}$
are the third
components of the isovector vector and axial currents,
$j^{\mathrm{em}}_{\alpha}$
is the electromagnetic current and
\begin{equation}
v^{S}_{\alpha}
=
\bar{s} \gamma_{\alpha} s
\;,
\qquad
a^{S}_{\alpha}
=
\bar{s} \gamma_{\alpha} \gamma_{5} s
\label{04}
\end{equation}
are the strange vector and axial currents. 
From Eq.(\ref{03})
it follows that using only the isotopic SU(2) invariance
of strong interactions we can connect the contributions of
the strange
currents to the cross sections of
the NC-induced processes with 
measurable quantities.
In Ref.\cite{ABGM96} we have considered
possibilities to extract information about
the strange axial and vector
form factors from the measurement
of the cross sections of NC elastic
and CC quasi-elastic
neutrino-nucleon scattering.

We will consider here the
\textbf{inclusive}
processes
\begin{equation}
\nu_{\mu}
\,
(\bar\nu_{\mu})
+
A
\to
\nu_{\mu}
\,
(\bar\nu_{\mu})
+
X
\;,
\label{05}
\end{equation}
where $A$
is a nucleus with isotopic spin $T$ equal to zero
(as $d$, $^{4}$He, $^{12}$C, etc.).
For a $T=0$ nucleus,
the interference 
between the isovector and isoscalar terms
of the hadronic neutral current
do not contribute to the
cross sections of the processes (\ref{05}).
In order to separate the contribution of the
axial strange current 
to the cross sections, 
let us consider the difference
\begin{equation}
\left(
\frac
{ \mathrm{d}\sigma }
{ \mathrm{d}Q^2 \mathrm{d}\nu }
\right)^{\mathrm{NC}}_{\nu}
-
\left(
\frac
{ \mathrm{d}\sigma }
{ \mathrm{d}Q^2 \mathrm{d}\nu }
\right)^{\mathrm{NC}}_{\bar\nu}
=
-
\frac{G_F^2}{2\pi}
\left( \frac{ M }{ p \cdot k } \right)^2
L_{5}^{\alpha\beta}(k,k')
\,
W^{\mathrm{NC};I}_{\alpha\beta}(p,q)
\;,
\label{06}
\end{equation}
where $p$ is the nucleus momentum,
$k$ and $k'$ are the  momenta of the initial and
final neutrinos (antineutrinos),
$ q = k-k' $,
$ Q^2 = - q^2 $,
\begin{equation}
L_{5}^{\alpha\beta}(k,k')
=
i
\,
\epsilon^{\alpha\beta\rho\sigma}
\,
k_{\rho}
\,
k'_{\sigma}
\;,
\label{061}
\end{equation}
and
the pseudotensor
$W^{\mathrm{NC};I}_{\alpha\beta}(p,q)$
is due to the interference of the
hadronic vector and axial currents.
We have
\begin{equation}
W^{\mathrm{NC};I}_{\alpha\beta}(p,q)
=
\left(
1
-
2 \sin^2 \theta_{W}
\right)
W^{V^{3};A^{3}}_{\alpha\beta}(p,q)
+
\sin^2 \theta_{W}
W^{V^{0};A^{S}}_{\alpha\beta}(p,q)
\;.
\label{07}
\end{equation}
The first term of this expression is due to
the interference of
$V^{3}$ and $A^{3}$.
From the isotopic
invariance of strong interactions it follows that
\begin{equation}
W^{V^{3};A^{3}}_{\alpha\beta}(p,q)
=
\frac{1}{2}
W^{\mathrm{CC};I}_{\alpha\beta}(p,q)
\;,
\label{08}
\end{equation}
where the pseudotensor
$W^{\mathrm{CC};I}_{\alpha\beta}(p,q)$
determines the difference
of the cross sections of the CC inclusive processes
\begin{equation}
\nu_{\mu}
\,
(\bar\nu_{\mu})
+
A
\to
\mu^{-}
\,
(\mu^{+})
+
X
\label{09}
\end{equation}
for relatively small energies
(up to the threshold of charm production).

The pseudotensor
$W^{V^{0};A^{S}}_{\alpha\beta}(p,q)$
in the Eq.(\ref{07})
is due to the interference of
the isoscalar part of electromagnetic
current and the strange axial current.
With the help
of Eqs.(\ref{06}) and (\ref{08}),
for the asymmetry
\begin{equation}
A
=
\frac
{ \displaystyle
\left(
\frac
{ \mathrm{d}\sigma }
{ \mathrm{d}Q^2 \mathrm{d}\nu }
\right)^{\mathrm{NC}}_{\nu}
-
\left(
\frac
{ \mathrm{d}\sigma }
{ \mathrm{d}Q^2 \mathrm{d}\nu }
\right)^{\mathrm{NC}}_{\bar\nu}
}
{ \displaystyle
\left(
\frac
{ \mathrm{d}\sigma }
{ \mathrm{d}Q^2 \mathrm{d}\nu }
\right)^{\mathrm{CC}}_{\nu}
-
\left(
\frac
{ \mathrm{d}\sigma }
{ \mathrm{d}Q^2 \mathrm{d}\nu }
\right)^{\mathrm{CC}}_{\bar\nu}
}
\label{10}
\end{equation}
we find the following expression
\begin{equation}
A
\,
|V_{ud}|^2
=
\frac{1}{2}
\left(
1
-
2 \sin^2 \theta_{W}
\right)
+
\frac{1}{2}
\,
\sin^2 \theta_{W}
\,
\frac
{ L_{5}^{\alpha\beta} \, W^{V^{0};A^{S}}_{\alpha\beta} }
{ L_{5}^{\alpha\beta} \, W^{V^{3};A^{3}}_{\alpha\beta} }
\;.
\label{11}
\end{equation}
The values of parameters
$|V_{ud}|$
and
$\sin^2 \theta_{W}$
is known with high accuracy
($ |V_{ud}| = 0.9736 \pm 0.0010 $
and
$ \sin^2 \theta_{W} = 0.226 \pm 0.004 $
\cite{PDG96}).
Thus,
a measurement of the asymmetry $A$ could allow to obtain
a model independent information about the matrix element of
the axial strange current. 

Let us notice that in
the deep inelastic region the last term of
Eq.(\ref{11})
is small
\cite{Paschos-Wirbel}
and the corresponding relation
for the total asymmetry
(the so-called Paschos-Wolfenstein relation
\cite{Paschos-Wolfenstein})
is used for the determination
of the value of
$\sin^2 \theta_{W}$.
We are proposing to measure
the asymmetry $A$ in the
intermediate energy region,
in which nuclei behave like bound states of
\textbf{nucleons}
and the contribution of the electromagnetic isoscalar current
is not suppressed
\cite{Donnelly92}.

\bigskip

It is a pleasure for us to
thank A. Molinari
for useful discussions
on the problem
of the isoscalar current.

\end{document}